\documentclass[letter]{IEEEtran}
\usepackage{amsmath,amsfonts}
\usepackage{algorithmic}
\usepackage{array}
\usepackage{textcomp}
\usepackage{stfloats}
\usepackage{url}
\usepackage{verbatim}
\usepackage{graphicx}
\usepackage{float}
\usepackage{subfigure}
\usepackage{amsmath}
\usepackage{amssymb}
\usepackage{caption2}
\usepackage{cite}
\usepackage{bm}
\usepackage[colorlinks,
            linkcolor=black,
            anchorcolor=black,
            citecolor=blue]{hyperref}
\def\BibTeX{{\rm B\kern-.05em{\sc i\kern-.025em b}\kern-.08em
    T\kern-.1667em\lower.7ex\hbox{E}\kern-.125emX}}
\usepackage{balance}
\begin{document}
\title{Power Allocation for Coordinated Multi-Point Aided ISAC Systems}

\author{Jianpeng Zou, Zhanfeng Zhong, Jintao Wang, Zheng Shi, Guanghua Yang and Shaodan Ma
\thanks{Jianpeng Zou, Zhanfeng Zhong, Jintao Wang, Zheng Shi, and Guanghua Yang are with the School of Intelligent Systems Science and Engineering, Jinan University, Zhuhai 519070, China (e-mails: z010222@stu2022.jnu.edu.cn; 1901542516@qq.com; wang.jintao@connect.um.edu.mo; zhengshi@jnu.edu.cn; ghyang@jnu.edu.cn).}
\thanks{Shaodan Ma is the State Key Laboratory of Internet of Things for Smart City, University of Macau, Macau 999078, China (e-mail: shaodanma@um.edu.mo).}
        }

\maketitle

\begin{abstract} 

In this letter, we investigate a coordinated multiple point (CoMP)-aided integrated sensing and communication (ISAC) system that supports multiple users and targets. Multiple base stations (BSs) employ a coordinated power allocation strategy to serve their associated single-antenna communication users (CUs) while utilizing the echo signals for joint radar target (RT) detection. The probability of detection (PoD) of the CoMP-ISAC system is then proposed for assessing the sensing performance. To maximize the sum rate while ensuring the PoD for each RT and adhering to the total transmit power budget across all BSs, we introduce an efficient power allocation strategy. Finally, simulation results are provided to validate the analytical findings, demonstrating that the proposed power allocation scheme effectively enhances the sum rate while satisfying the sensing requirements.

\end{abstract}


\section{Introduction}
\IEEEPARstart{D}{ue} to the scarcity of spectrum resources and the shift toward higher frequency bands, communications and sensing are becoming increasingly similar in terms of signal processing, channel characteristics, and hardware structures  \cite{10418473}. Consequently, the integration of communication and sensing plays a vital role in improving the efficiency of wireless resources.
The integrated sensing and communication (ISAC) system is expected to provide wireless communication and radar sensing functions, which promotes the realization of new intelligent applications and services in the future sixth-generation (6G) communication systems, such as unmanned aerial vehicle communication and sensing, vehicle networking, and smart cities \cite{zhang2021enabling}.
%


ISAC has been widely studied on waveform design, performance analysis and signal processing, etc \cite{xiao2022waveform,ouyang2022performance,guo2023joint}. For example, a novel full-duplex (FD) ISAC waveform design scheme has been proposed in \cite{xiao2022waveform}, wherein a single ISAC node aims to simultaneously sense a radar target while communicating with a receiver. The analysis conducted in \cite{ouyang2022performance} explored the expressions for outage probability, ergodic communication rate, and sensing rate within the uplink non-orthogonal multiple access (NOMA)-ISAC system, also revealing the diversity order and the high signal-to-noise ratio (SNR) slope. Furthermore, the joint design problem encompassing beamforming, power allocation, and signal processing in a reconfigurable intelligent surface (RIS)-assisted ISAC system was examined in \cite{guo2023joint}, with optimization challenges addressed through optimal minimization and penalty dual decomposition techniques.
However, the aforementioned works primarily focus on single-cell scenarios, which significantly limits their applicability.

With advancements in coordinated multi-point transmission (CoMP) \cite{gesbert2010multi}, cloud radio access networks \cite{wu2015cloud}, and cell-free multiple-input multiple-output (MIMO) systems \cite{bjornson2020scalable}, multipoint cooperative ISAC emerges as a promising and natural architecture for enhancing performance. Network-level ISAC, which relies on multi-cell cooperation, can effectively expand both communication and sensing coverage while providing an additional degree of freedom (DoF) to achieve greater integration gains between communication and sensing \cite{meng2024cooperative}.
For instance, authors in \cite{xu2023integrated} proposed a coordinated cellular network-supported multi-static radar architecture for ISAC systems, enabling spatial separation of signal transmission and radar echo reception to mitigate self-interference while optimizing beamforming strategies. Authors in \cite{huang2024edge} introduced an edge intelligence-oriented ISAC approach, where multiple ISAC stations collaborate to sense targets and offload data to a powerful edge server for model training, utilizing a hierarchical block coordinate descent algorithm for joint optimization. 
Additionally, \cite{zhang2023joint} developed an interference model for ISAC in dense cellular networks (DCNs) and addressed multidimensional resource allocation by decoupling the optimization problem into subproblems, employing a greedy genetic subband allocation scheme to reduce interference and using geometric programming for transmission power control.
Unfortunately, research on coordinated ISAC transmission primarily emphasizes system optimization, which often results in high complexity and a lack of comprehensive performance evaluation.

In this letter, we investigate a CoMP-ISAC network with multiple base stations (BSs) in multiple cells employing coordinated power allocation to serve their associated single-antenna user equipment (CUs) while utilizing echo signals for joint radar target (RT) detection. We provide a quantitative measure of the sensing performance by deriving the probability of detection (PoD) of the CoMP-ISAC system.
A power allocation strategy aimed at maximizing the communication sum rate while adhering to constraints on total transmit power and PoD is proposed.  Numerical results validate the theoretical analysis and demonstrate that the proposed power allocation scheme effectively enhances the communication sum rate of the system.



\section{System Model}\label{sec:system_model}

\begin{figure}[t]
\centering
\includegraphics[width=0.45\textwidth]{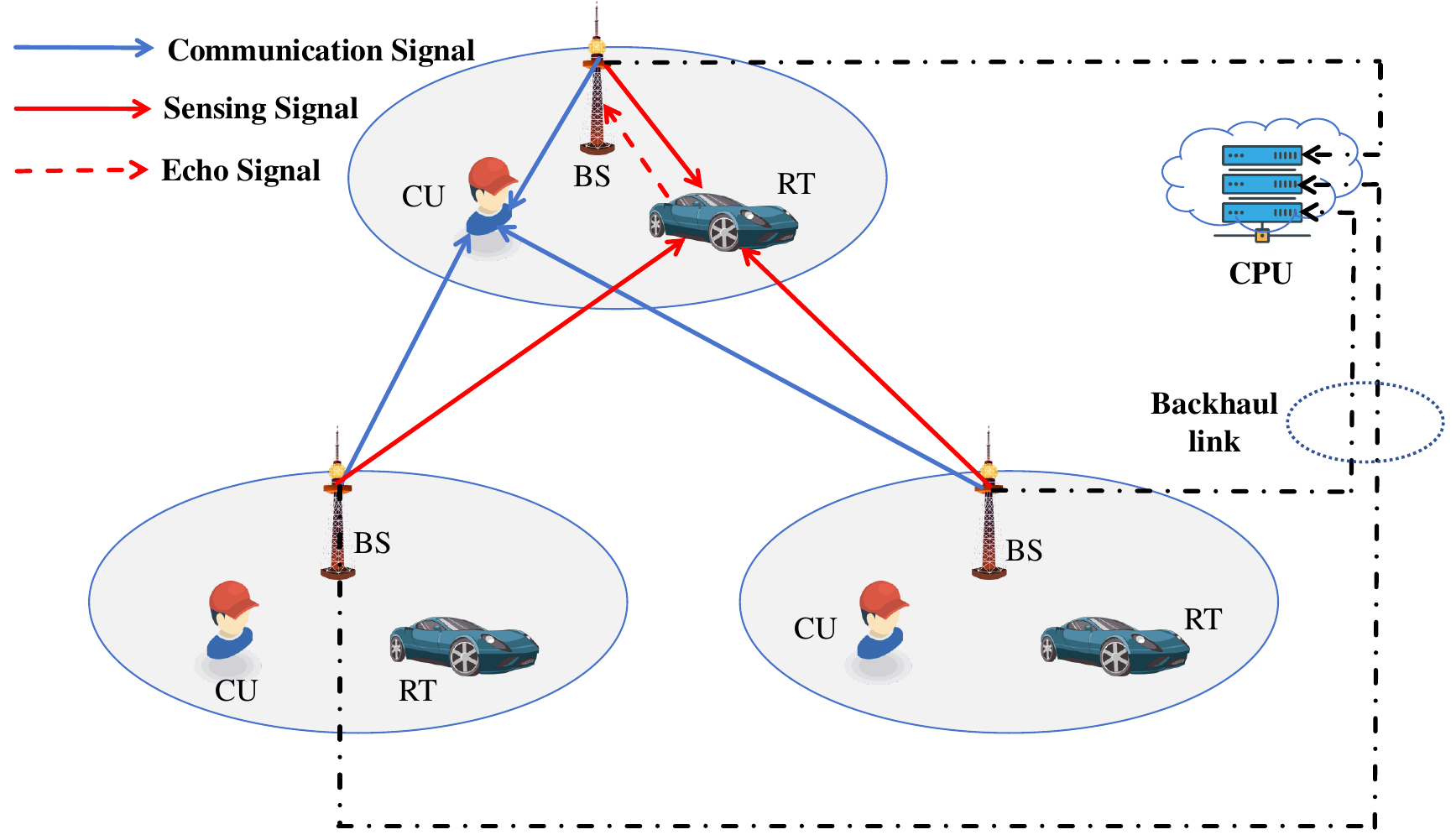}
\caption{\label{fig:model} Illustration of a CoMP-ISAC network comprising $L=3$ cells.}
\end{figure}

In this paper, we consider a CoMP-ISAC network consisting of $L$ cells, each featuring one BS, one CU, and one RT. Each BS is equipped with independent transmitting and receiving antennas to provide communication services to the user while simultaneously detecting the target within its cell.
For cell $C_i$, the corresponding BS is denoted as $B_i$, located at the center of a fixed circular coverage area with a radius $r$. The BS $B_i$ serves its user $U_i$ and senses its target $T_i$, with both $U_i$ and $T_i$ positioned within the coverage region of  $B_i$ \cite{jia2024interference}. A simple illustration of a CoMP-ISAC network comprising $L=3$ cells is depicted in Fig.~\ref{fig:model}. 


Suppose the BS $B_i$ broadcasts a modulated signal $x_{i}$ to sense the target $T_i$ and communicate with the $U_i$ simultaneously. The received signal at the user $U_i$ is represented as 
\begin{equation}
\label{comm_signal}
y_{c,i} = \sqrt{P_i}h_{i,i}^{c}x_i + \sum_{l\neq i}^{L}\sqrt{P_l}h_{l,i}^{c}x_l + n_{c,i} ,
\end{equation} 
where $P_i$ denotes the transmit power of BS $B_i$ and the second term represents the intercell interference.
$h_{l,i}^c\sim \mathcal{CN}(0,\sigma_{l,i}^2)$ denotes the channel coefficients from BS $B_l$ to the communication user $U_i$, and $n_{c,i} \sim \mathcal{CN}(0,\sigma_{c,i}^2)$ refers to the additive Gaussian white noise (AWGN).
Thus, the communication rate at the user $U_i$ is derived as
\begin{align}
    R_{i}=\log_2\left( 1+\frac{P_i|h^c_{i,i}|^2}{\sum_{l\neq i}^{L}P_l|h^c_{l,i}|^2 + \sigma_{c,i}^2} \right),
\end{align}
Therefore, the sum rate of all communication users across $L$ cells is denoted as
\begin{align}
    R_{sum} = \sum_{i=1}^{L} R_i.
\end{align}

The echo signal received at the BS $B_i$ is expressed as \cite{cheng2024optimal}
\begin{equation}
\label{sen_signal}
y_{s,i} = \sum_{l= 1}^{L}\sqrt{P_l}h_{l,i}^{s}x_l + n_{s,i},
\end{equation}
where $h_{l,i}^{s}$ denotes the two-round channel coefficients associated with the $B_l$-to-$T_i$-to-$B_i$ links, and $n_{s,i} \sim \mathcal{CN}(0,\sigma_{s,i}^2)$ represents the AWGN at the BS $B_i$.

\section{Probability of Detection} \label{sec:performance_metrice}
In this section, we derive the probability of detection (PoD) of the proposed CoMP-ISAC network, thereby providing a quantitative measure of the sensing performance.

Taking $N$ observation samples, the received signal $\mathbf{y}_{s,i}=[y_{s,i}^{1},y_{s,i}^{2},\cdots,y_{s,i}^{N}]^T$ can be rewritten as
\begin{equation}
\label{sensing:signal}
\begin{aligned}
\mathbf{y}_{s,i} = \mathbf{X}\mathbf{h}_s + \mathbf{n}_{s,i} ,
\end{aligned}
\end{equation}
where  $\mathbf{X} = \left(\sqrt{P_1}\mathbf{x}_1,\sqrt{P_2}\mathbf{x}_2,\cdots,\sqrt{P_L}\mathbf{x}_L\right)$, with $\mathbf{x}_l = \left(x_{l}^{1},x_{l}^{2},\cdots,x_{l}^{N}\right)^T$ and $\mathbf{n}_{s,i}=\left(n_{s,i}^{1},n_{s,i}^{2},\cdots,n_{s,i}^{N}\right)^T$.  $\mathbf{h}_s$ represents the concatenated two-round channels  regarding the BS $B_i$ with $\mathbf{h}_s = \left(h_{1,i}^s,h_{2,i}^s,\cdots,h_{L,i}^s\right)^T$.

To start with, we define two hypotheses for target detection, i.e., $\mathcal{H}_0$ when the target does not exist and $\mathcal{H}_1$ when the target exists. Then, based on \eqref{sensing:signal}, the received signals at the BS $B_i$ can be expressed as
%
\begin{align} \label{binary hypothesis}
\left\{
\begin{array}{l}
\mathcal{H}_0:\mathbf{y}_{s,i} = \mathbf{n}_{s,i}, \\
\mathcal{H}_1:\mathbf{y}_{s,i} = \mathbf{X} \mathbf{h}_{s} + \mathbf{n}_{s,i} .
\end{array}
\right.
\end{align}

Next, we adopt the methodology of the likelihood ratio test (LRT) for target detection. 
We first estimate $\mathbf{h}_{s}$ by an estimate in the LRT function, which leads to a new test function that becomes the generalized LRT (GLRT) \cite{an2023fundamental}. Therefore, the maximum likelihood estimate of $\mathbf{h}_{s}$ can be obtained by
\begin{equation}
\label{h_hat}
\begin{aligned}
\widehat{\mathbf{h}}_s =  (\mathbf{X}^H\mathbf{X})^{-1} \mathbf{X}^H \mathbf{y}_{s,i}.
\end{aligned}
\end{equation}
Based on \eqref{binary hypothesis} and \eqref{h_hat}, the logarithmic LRT function is given by
\begin{align}\label{P_FA}
    \ln\Xi(\mathbf{y}_{s,i})=\frac{1}{\sigma_{s,i}^2} \left( \mathbf{y}_{s,i}^H \mathbf{X}(\mathbf{X}^H\mathbf{X})^{-1}\mathbf{X}^H \mathbf{y}_{s,i} \right). 
\end{align}

\subsection{Probability of False Alarm}
Given the threshold $\delta$, the probability of false alarm (PFA) undr the hypothesis $\mathcal{H}_0$ is derived as
\begin{equation}
\label{P_{FA}}
\begin{aligned}
\mathcal{P}_{FA}^{i} &= \Pr \left\{ \frac{1}{\sigma_{s,i}^2} \mathbf{y}_{s,i}^{H} \mathbf{X}(\mathbf{X}^H\mathbf{X})^{-1}\mathbf{X}^H \mathbf{y}_{s,i} \geq \delta \Big{|} \mathcal{H}_0 \right\}.
\end{aligned}
\end{equation}
Letting $\mathbf{P} \triangleq \mathbf{X}(\mathbf{X}^H\mathbf{X})^{-1}\mathbf{X}^H$, $\mathbf{P}$ denotes the orthogonal projection matrix corresponding to $\mathbf{X}$, satisfying $\mathbf{P}^2=\mathbf{P}=\mathbf{P}^H$. 
Given that the rank of $\mathbf{P}$ is $\mathrm{rank}(\mathbf{P}) =L$, there are $L$ eigenvalues of $\mathbf{P}$ equal to 1. Therefore, $\mathbf{P}$ can be expressed as $\mathbf{P}=\mathbf{U}^H{\bm{\Lambda}}\mathbf{U}$ using eigenvalue decomposition, where ${\bm{\Lambda}}$ is a diagonal matrix with $L$ eigenvalues equal to 1, and all other diagonal entries are zero. Then, we can rewrite \eqref{P_FA} as 
\begin{subequations}
\begin{align}
\mathcal{P}_{FA}^{i} &= \Pr \left\{  \frac{1}{\sigma_{s,i}^2} \mathbf{n}_{s,i}^H \mathbf{U}^H{\bm{\Lambda}}\mathbf{U} \mathbf{n}_{s,i} \geq  \delta \right\}, \\
& = \frac{\Gamma(L,\delta)}{\Gamma(L)},
\end{align}
\end{subequations}
where $\Gamma(\cdot,\cdot)$ is the upper incomplete Gamma function.
The second equality holds since $\mathbf{n}_{s,i} \sim \mathcal{CN}(0,\sigma_{s,i}^2\mathbf{I})$ and $\mathbf{z}=\mathbf{U}\mathbf{n}_{s,i} / \sigma_{s,i}$. Consequencely, $\mathbf{z}^H\Lambda\mathbf{z}$ follows a central chi-squared distribution with $2L$ degree of freedom.

\subsection{Probability of Detection}
Similarly, the probability of detection (PoD) under the hypothesis $\mathcal{H}_1$ can be written as
\begin{align}
P_{D}^{i} = \Pr \left\{ \frac{1}{\sigma_{s,i}^2} \mathbf{y}_{s,i}^H \mathbf{X}(\mathbf{X}^H\mathbf{X})^{-1}\mathbf{X}^H \mathbf{y}_{s,i} \geq \delta \Big{|} \mathcal{H}_1 \right\}, 
\end{align}
Substituting \eqref{binary hypothesis} and denoting $\mathbf{\tilde{n}} = \mathbf{U}\mathbf{n}_{s,i}$ and $\mathbf{q}=\mathbf{UX}\mathbf{h}_s$, the  PoD $P_{D}^{i}$ can be transformed as 
\begin{equation}
\label{P_{D}}
\begin{aligned}
P_{D}^{i} & = \Pr \left\{ \frac{1}{\sigma_{s,i}^2}(\Lambda\mathbf{q}+\Lambda\mathbf{\tilde{n}})^H (\Lambda\mathbf{q}+\Lambda\mathbf{\tilde{n}}) \geq \delta \Big{|} \mathcal{H}_1 \right\}, \\
&\overset{(a)}{=} \Pr \left\{ (\mathbf{q}_L+\mathbf{\tilde{n}}_L)^H (\mathbf{q}_L+\mathbf{\tilde{n}}_L) \geq \delta \Big{|} \mathcal{H}_1 \right\}, \\
&\overset{(b)}{=} Q_L \left(\sqrt{2\mathbf{q}_L^H\mathbf{q}_L},\sqrt{2\delta}\right), \\
&= Q_L \left(\sqrt{\frac{2}{\sigma_{s,i}^2}\mathbf{h}_s^H\mathbf{X}^H\mathbf{X}\mathbf{h}_s},\sqrt{2\delta}\right), \\
&\overset{N\gg 1}{\simeq} Q_L \left(\sqrt{\frac{2}{\sigma_{s,i}^2}\sum_{l=1}^{L}P_LN|h_{l,i}^s|^2},\sqrt{2\delta}\right),
\end{aligned}
\end{equation}
where $Q_L(\cdot)$ is the Marcum-Q function of order $L$.
Step (a) holds by defining $\mathbf{q}_L=\left[ \frac{1}{\sigma_{s,i}} \Lambda \mathbf{q} \right]_{1:L}$ and $\mathbf{\tilde{n}}_L=\left[ \frac{1}{\sigma_{s,i}} \Lambda \mathbf{\tilde{n}} \right]_{1:L}$.
and step (b) holds since $2(\mathbf{q}_L+\mathbf{\tilde{n}}_L)^H(\mathbf{q}_L+\mathbf{\tilde{n}}_L)$ is a non-central chi-squared distribution with degree of freedom $2L$   \cite{krishnan1967noncentral}.

\section{Power Allocation Strategy}\label{sec:optimion}
In this section, we focus on maximizing the transmission sum rate of communication by optimizing the transmit power at each BS, while ensuring the PoD and adhering to the total transmit power budget.

The sum rate maximization problem is then formulated as
\begin{subequations}
\label{opt:problem}
\begin{align}
	\mathop{\max}\limits_{P_{1},\cdots,P_{L}} & \quad R_{sum} \label{opt:obj} \\
	s.t.
		& \quad \sum\nolimits_{l=1}^{L}P_l \leq P_{th}, \label{opt:power} \\
            & \quad R_{l} \geq R_{l,th}^c , l\in [1,\cdots,L], \label{opt:comm} \\
		& \quad P_{D}^l  \geq \xi_{S,th}^l , l\in [1,\cdots,L]\label{opt:PD} ,
\end{align}
\end{subequations}
where $P_{th}$, $R_{l,th}^c$ and $\xi_{S,th}^l$ denote the threshold of total transmit power, the minimum required communication rate for the $l$-th CU and the minimum required PoD for the $l$-th target, respectively. 
However, the fractional form of the objective function, combined with the non-convex feasible set, significantly complicates the optimization of \eqref{opt:problem}.


By defining $\rho_i=|h^c_{i,i}|^2$ and $\rho_l=|h^c_{l,i}|^2$, the original objective function \eqref{opt:obj} can be transformed as
\begin{equation}
\label{second_obj}
\begin{aligned}
R_{sum}\!=\!	\sum_{i=1}^{L} \left[ \log_2\left( \sum_{l=1}^{L}P_l\rho_l \!+\! \sigma_{c,i}^2 \right) \!-\! \log_2\left(\sum_{l\neq i}^{L}P_l\rho_l \!+\! \sigma_{c,i}^2 \right) \right] , 
\end{aligned}
\end{equation}
We can observe that \eqref{second_obj} is still challenging to solve due to the non-convexity of the negative logarithmic functions, i.e., $-\log_2\left(\sum_{l\neq i}^{L}P_l\rho_l + \sigma_{c,i}^2 \right)$. To this end, according to \cite{liu2023non}, we adopt the following function to tackle this challenge 
\begin{equation}
\label{-lnx}
\begin{aligned}
	g(t) = -tx+\ln t + 1, \quad t>0 , 
\end{aligned}
\end{equation}
It can be inferred from \eqref{-lnx} that if and only if $t^* = 1/x$, the maximum of $g(t)$ can be obtained, i.e., $-\ln x = \mathop{\max}\limits_{t>0} g(t)$. In this way, \eqref{second_obj} can be reformulated as
\begin{equation}
\label{second_opt}
\begin{aligned}
	R_{sum}= \sum_{i=1}^{L} \left[ \log_2\left( \sum_{l=1}^{L}P_l\rho_l + \sigma_{c,i}^2 \right)+ \mathop{\max}\limits_{t_{i}} \frac{g(t_i)}{\ln 2} \right] , 
\end{aligned}
\end{equation}
where $g(t_i) = -t_i \left( \sum_{l\neq i}^{L}P_l\rho_l + \sigma_{c,i}^2 \right) + \ln t_i +1$.

Since $R_l$ is a monotonically increasing function of communication SNR, \eqref{opt:comm} can be further relaxed as
\begin{equation}
\label{rate_to_SNR}
\begin{aligned}
	\frac{P_i|h_{i,i}^c|^2}{\sum_{l\neq i}^{L}P_l|h_{l,i}^c|^2+\sigma_{c,i}^2} \geq \zeta_{l,th}^c, l\in[1,\cdots,L] ,
\end{aligned}
\end{equation}
where $\zeta_{l,th}^c$ is the corresponding threshold.
Similarly, $P_D^l$ is a monotonically increasing function of sensing SNR and \eqref{opt:PD} can be further relaxed as
\begin{equation}
\label{PD_to_SNR}
\begin{aligned}
	\frac{\sum_{l=1}^{L}P_l|h_{l,i}^s|^2}{\sigma_{s,i}^2} \geq \zeta_{th}^l, l\in[1,\cdots,L] ,
\end{aligned}
\end{equation}
where $\zeta_{th}^l$ is the corresponding threshold.
After these transformations, the sum rate maximization problem \eqref{opt:problem} can be rewritten as
\begin{equation}
\label{frist_opt pro}
\begin{aligned}
\mathop{\max}\limits_{P_{1},\cdots,P_{L}} & \quad \eqref{second_opt} \\
	s.t.
		& \quad t_i>0, \quad \eqref{opt:power} 
            \quad \eqref{rate_to_SNR}
		  \quad \eqref{PD_to_SNR} ,
\end{aligned}
\end{equation}
%
 Note that the optimization problem \eqref{frist_opt pro} is convex and can be efficiently solved by the popular CVX toolbox.

\section{Numerical Results}\label{sec:NumRes}
\begin{figure*}[htbp]
	\centering
	\subfigure[] {\includegraphics[width=.32\textwidth]{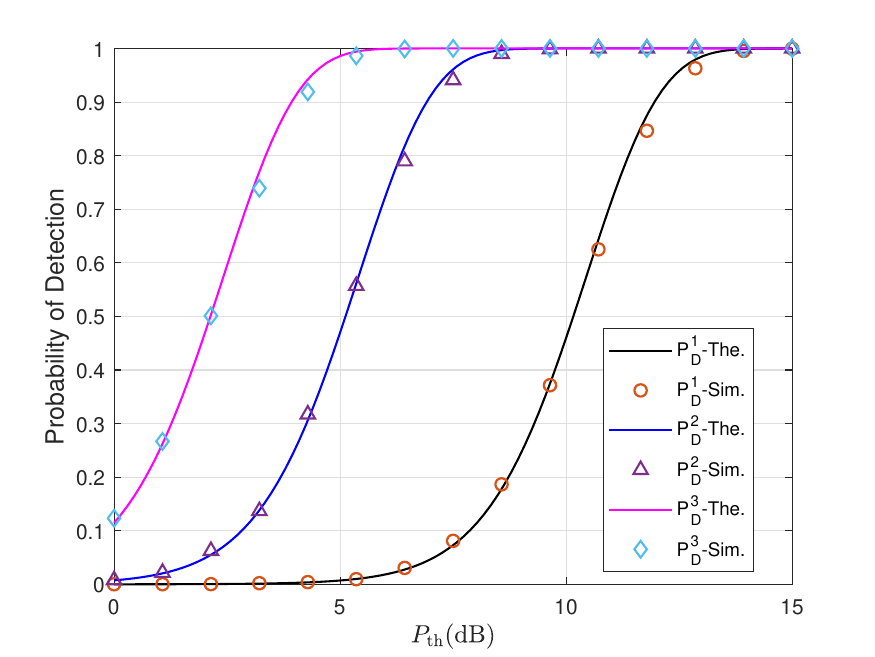}}
    \hspace{-6pt}
	\subfigure[] {\includegraphics[width=.32\textwidth]{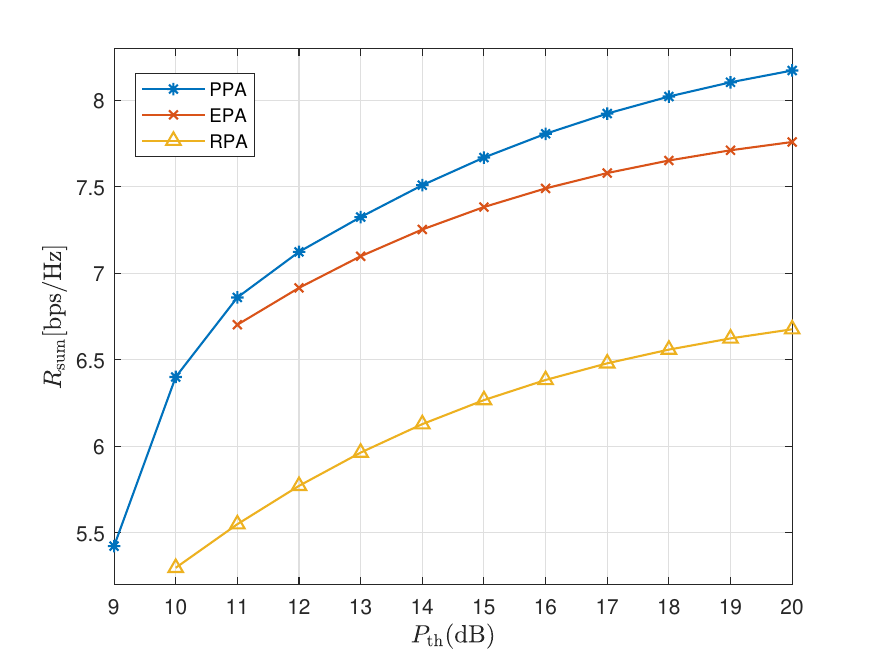}}
    \hspace{-6pt}
	\subfigure[] {\includegraphics[width=.32\textwidth]{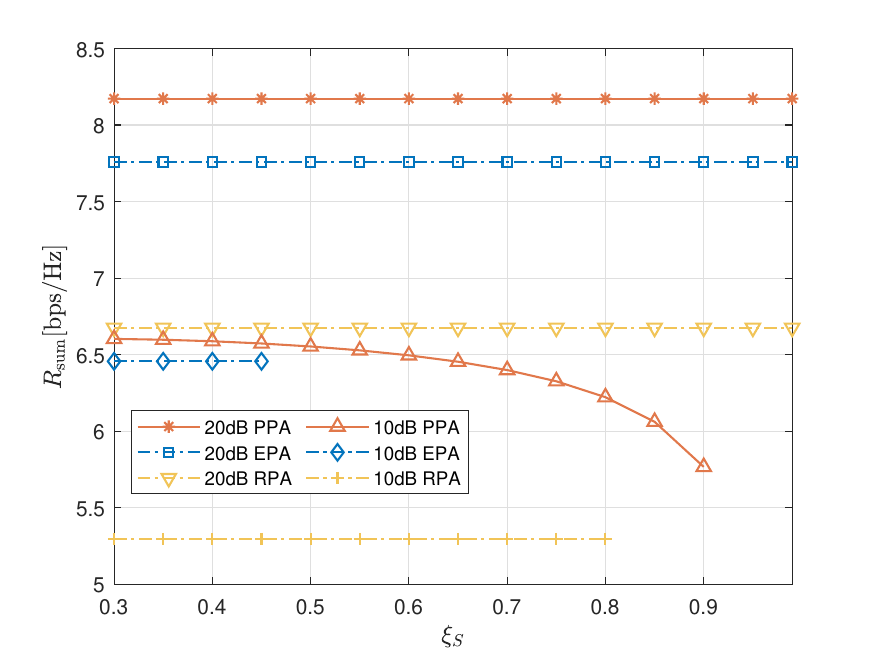}}
	\caption{(a) Probability of detection versus varying power budgets; (b) sum rate versus varying power budgets; (c) sum rate versus varying thresholds of the probability of detection.}
    \vspace{-12pt}
	\label{fig}
\end{figure*}

In this section, we present simulation results for verification and discussion. For illustration, we assume that $L=3$,  $N=100$, $\mathcal{P}_{\mathrm{FA}}^i=10^{-6}$, $\sigma_{c,i}^2=1$ dB, $\sigma_{s,i}^2=15$ dB, $\xi_{S,th}^i=\xi_S=0.7$, $\xi_{C,th}^i=\xi_C=1$ bps/Hz for $\forall i$. The proposed power allocation scheme in this work is labeled as PPA. The following benchmarks are used for comparison: the equal power allocation scheme (EPA) and the random power allocation scheme within the feasible region, with a random seed of 2 (RPA).
The theoretical and simulated values of PoD in the CoMP-ISAC system are plotted in Fig. \ref{fig} (a). The simulation results are obtained from 10,000 independent experiments. It can be observed that the theoretical values align well with the simulated values. 

Fig.~\ref{fig} (b) compares the communication sum rate performance versus the power budget $P_{th}$ for different PA schemes. Within the feasible region, the communication sum rate increases as the power budget increases.  In Fig.~\ref{fig} (b), PPA achieves the maximum communication sum rate for the same power budget, outperforming the other benchmarks.
At $P_{th}=9$ dB, only the proposed scheme meets both communication and sensing performance requirements.  As $P_{th}$ increases, the gap between the proposed scheme and the other benchmarks widens.
Notably, under $P_{th}=15$ dB, the sum rate $R_{sum}$ for the proposed scheme is roughly equivalent to that under $P_{th}=19$ dB for the EPA scheme, indicating a performance gain of 4 dB. Therefore, the proposed scheme demonstrates superior performance compared to the other benchmarks in both low and high-power budget regimes.

Fig. \ref{fig} (c) depicts the impact of varying the values of $\xi_{S}$ and $P_{th}$.
As $\xi_{S}$ increases, the sum rate $R_{sum}$ for other benchmarks remains constant until they reach the region of infeasible solutions. In contrast, rather than yielding an infeasible solution while maintaining the same $R_{sum}$, the proposed PPA adapts to meet the increasing sensing performance requirements, albeit at the cost of a reduced communication sum rate. Under $P_{th}=10$ dB, the $R_{sum}$ for the PPA gradually decreases but continues to outperform the other benchmarks as $\xi_S$ rises. 
When transitioning from a less stringent requirement of $\xi_S=0.3$ to a stricter requirement of $\xi_S=0.8$, the RPA scheme achieves a lower communication sum rate than the proposed scheme. In contrast, the proposed scheme maintains a more optimized communication sum rate while achieving better sensing performance, highlighting its advantages.
When $P_{th}$ increases to 20 dB, all power allocation schemes meet the sensing performance requirement of $\xi_S=0.99$ due to the higher power budget. However, among all the power allocation schemes depicted in Fig.~\ref{fig} (c), the proposed scheme achieves the best communication sum rate.

\section{Conclusion}\label{sec:Conclusion}
In this letter, we investigated an ISAC system within a CoMP network, where BSs can simultaneously perform communication and target sensing. To evaluate the sensing performance, we derived the expression for the PoD. Furthermore, we proposed an efficient power allocation scheme to maximize the communication sum rate while adhering to constraints on total transmit power and PoD. Finally, simulations confirmed that the proposed power allocation schemes effectively improve the sum rate of the system while satisfying the sensing requirements.

\bibliographystyle{IEEEtran}
\bibliography{IEEEabrv,main}

\begin{thebibliography}{10}
\providecommand{\url}[1]{#1}
\csname url@samestyle\endcsname
\providecommand{\newblock}{\relax}
\providecommand{\bibinfo}[2]{#2}
\providecommand{\BIBentrySTDinterwordspacing}{\spaceskip=0pt\relax}
\providecommand{\BIBentryALTinterwordstretchfactor}{4}
\providecommand{\BIBentryALTinterwordspacing}{\spaceskip=\fontdimen2\font plus
\BIBentryALTinterwordstretchfactor\fontdimen3\font minus \fontdimen4\font\relax}
\providecommand{\BIBforeignlanguage}[2]{{%
\expandafter\ifx\csname l@#1\endcsname\relax
\typeout{** WARNING: IEEEtran.bst: No hyphenation pattern has been}%
\typeout{** loaded for the language `#1'. Using the pattern for}%
\typeout{** the default language instead.}%
\else
\language=\csname l@#1\endcsname
\fi
#2}}
\providecommand{\BIBdecl}{\relax}
\BIBdecl

\bibitem{10418473}
S.~Lu, F.~Liu, Y.~Li, K.~Zhang, H.~Huang, J.~Zou, X.~Li, Y.~Dong, F.~Dong, J.~Zhu, Y.~Xiong, W.~Yuan, Y.~Cui, and L.~Hanzo, ``Integrated sensing and communications: Recent advances and ten open challenges,'' \emph{IEEE Internet Things J.}, vol.~11, no.~11, pp. 19\,094--19\,120, 2024.

\bibitem{zhang2021enabling}
J.~A. Zhang, M.~L. Rahman, K.~Wu, X.~Huang, Y.~J. Guo, S.~Chen, and J.~Yuan, ``Enabling joint communication and radar sensing in mobile networks—{A} survey,'' \emph{IEEE Commun. Surveys Tuts.}, vol.~24, no.~1, pp. 306--345, 1st Quart. 2022.

\bibitem{xiao2022waveform}
Z.~Xiao and Y.~Zeng, ``Waveform design and performance analysis for full-duplex integrated sensing and communication,'' \emph{IEEE J. Sel. Areas Commun.}, vol.~40, no.~6, pp. 1823--1837, Jun. 2022.

\bibitem{ouyang2022performance}
C.~Ouyang, Y.~Liu, and H.~Yang, ``On the performance of uplink {ISAC} systems,'' \emph{IEEE Commun. Lett.}, vol.~26, no.~8, pp. 1769--1773, Aug. 2022.

\bibitem{guo2023joint}
Y.~Guo, Y.~Liu, Q.~Wu, X.~Li, and Q.~Shi, ``Joint beamforming and power allocation for ris aided full-duplex integrated sensing and uplink communication system,'' \emph{IEEE Trans. Wireless Commun.}, vol.~23, no.~5, pp. 4627--4642, May 2024.

\bibitem{gesbert2010multi}
D.~Gesbert, S.~Hanly, H.~Huang, S.~S. Shitz, O.~Simeone, and W.~Yu, ``Multi-cell {MIMO} cooperative networks: {A} new look at interference,'' \emph{IEEE J. Sel. Areas commun.}, vol.~28, no.~9, pp. 1380--1408, Dec. 2010.

\bibitem{wu2015cloud}
J.~Wu, Z.~Zhang, Y.~Hong, and Y.~Wen, ``Cloud radio access network {(C-RAN)}: a primer,'' \emph{IEEE Netw.}, vol.~29, no.~1, pp. 35--41, Jan./Feb. 2015.

\bibitem{bjornson2020scalable}
E.~Bj{\"o}rnson and L.~Sanguinetti, ``Scalable cell-free massive {MIMO} systems,'' \emph{IEEE Trans. Commun.}, vol.~68, no.~7, pp. 4247--4261, Jul. 2020.

\bibitem{meng2024cooperative}
K.~Meng, C.~Masouros, A.~P. Petropulu, and L.~Hanzo, ``Cooperative {ISAC} networks: Opportunities and challenges,'' \emph{IEEE Wireless Communications.}, 2024.

\bibitem{xu2023integrated}
D.~Xu, C.~Liu, S.~Song, and D.~W.~K. Ng, ``Integrated sensing and communication in coordinated cellular networks,'' in \emph{Proc. IEEE Stat. Signal Process. Workshop (SSP)}.\hskip 1em plus 0.5em minus 0.4em\relax IEEE, Jul 2023, pp. 90--94.

\bibitem{huang2024edge}
N.~Huang, H.~Dong, C.~Dou, Y.~Wu, L.~Qian, S.~Ma, and R.~Lu, ``Edge intelligence oriented integrated sensing and communication: A multi-cell cooperative approach,'' \emph{IEEE Trans. Veh. Technol.}, vol.~73, no.~6, pp. 8810--8824, Jun. 2024.

\bibitem{zhang2023joint}
J.~Zhang, Z.~Fei, X.~Wang, P.~Liu, J.~Huang, and Z.~Zheng, ``Joint resource allocation and user association for multi-cell integrated sensing and communication systems,'' \emph{EURASIP J. Wireless Commun. Networking}, vol. 2023, no.~1, p.~64, 2023.

\bibitem{jia2024interference}
C.~Jia, Z.~Zhao, L.~Sun, and T.~Q. Quek, ``An interference cancellation scheme of integrated sensing and communications in wireless networks,'' \emph{IEEE Wireless Commun. Lett.}, vol.~13, no.~12, pp. 3429--3433, Dec. 2024.

\bibitem{cheng2024optimal}
G.~Cheng, Y.~Fang, J.~Xu, and D.~W.~K. Ng, ``Optimal coordinated transmit beamforming for networked integrated sensing and communications,'' \emph{IEEE Trans. Wireless Commun.}, vol.~23, no.~8, pp. 8200--8214, 2024.

\bibitem{an2023fundamental}
J.~An, H.~Li, D.~W.~K. Ng, and C.~Yuen, ``Fundamental detection probability vs. achievable rate tradeoff in integrated sensing and communication systems,'' \emph{IEEE Trans. Wireless Commun.}, vol.~22, no.~12, pp. 9835--9853, Dec. 2023.

\bibitem{krishnan1967noncentral}
M.~Krishnan, ``The noncentral bivariate chi distribution,'' \emph{SIAM Review}, vol.~9, no.~4, pp. 708--714, Oct. 1967.

\bibitem{liu2023non}
M.~Liu, M.~Yang, F.~Wei, H.~Li, Z.~Zhang, A.~Nallanathan, and L.~Hanzo, ``A non-orthogonal uplink/downlink {IoT} solution for next-generation {ISAC} systems,'' \emph{IEEE Internet Things J.}, vol.~11, no.~5, pp. 8224--8239, Mar 2023.

\end{thebibliography}

\end{document}